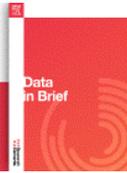

# ARTICLE INFORMATION

**Authors**

*Shijin Chen[a], Zeyi Liu[b], Xiao He[b], Dongliang Zou[a], Donghua Zhou[b,c]*

**Affiliations**

a MCC5 Group Shanghai Co. LTD, 201900, ShangHai, China.

b Department of Automation, Tsinghua University, 100084, Beijing, China.

c College of Electrical Engineering and Automation, Shandong University of Science and Technology, Qingdao 266590, China.

**Corresponding author's email address and Twitter handle**

*hexiao@tsinghua.edu.cn*

**Abstract**

The gearbox is a critical component of electromechanical systems. The occurrence of multiple faults can significantly impact system accuracy and service life. The vibration signal of the gearbox is an effective indicator of its operational status and fault information. However, gearboxes in real industrial settings often operate under variable working conditions, such as varying speeds and loads. It is a significant and challenging research area to complete the gearbox fault diagnosis procedure under varying operating conditions using vibration signals. This data article presents vibration datasets collected from a gearbox exhibiting various fault degrees of severity and fault types, operating under diverse speed and load conditions. These faults are manually implanted into the gears or bearings through precise machining processes, which include health, missing teeth, wear, pitting, root cracks, and broken teeth. Several kinds of actual compound faults are also encompassed. The development of these datasets facilitates testing the effectiveness and reliability of newly developed fault diagnosis methods.

# SPECIFICATIONS TABLE

| Subject | Mechanical engineering |
|---|---|
| Specific subject area | Vibration, machine condition monitoring, fault diagnosis |

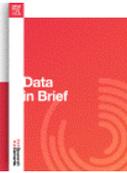

**Data in Brief** — Open access

Article template

| Type of data | Datasets in ".csv" format |
|---|---|
| Data collection | Two three-axis acceleration sensors were used to collect the three-axis vibration acceleration signals of the motor drive end and the gearbox intermediate shaft respectively. The photoelectric sensor was used to collect the key phase signal (speed data) of the motor output shaft. The torque sensor was used to collect the torque data of the gearbox input shaft. |
| Data source location | MCC5 Group Shanghai Co. LTD<br><br>Department of Automation, Tsinghua University |
| Data accessibility | Repository name: Multi-mode Fault Diagnosis Datasets of Gearbox Under Variable Working Conditions<br><br>Data identification number: 10.17632/p92gj2732w.1<br><br>Direct URL to data: https://data.mendeley.com/datasets/p92gj2732w/1<br><br>Instructions for accessing these data: |
| Related research article | Z. Liu, C. Li and X. He, Evidential ensemble preference-guided learning approach for real-time multimode fault diagnosis, *IEEE Transactions on Industrial Informatics*, doi: 10.1109/TII.2023.3332112. |

## VALUE OF THE DATA

1) The data is collected from gearboxes operating under variable working conditions, including time-varying speed and load. The dataset includes vibration signals, speed signals, and torque signals, and encompasses a variety of fault types, including multiple single gear faults and multiple bearing-gear compound faults, as well as different degrees of severity. As depicted in Table 1, our dataset offers advantages over existing representative datasets.
2) Our dataset distinguishes itself from existing literature by incorporating data from more complex variable working conditions, a broader range of fault types and severity degrees, and multiple signals.
3) This dataset enables the study of the time and frequency characteristics of gearbox fault signals under variable working conditions.
4) The dataset can be utilized to evaluate the effectiveness of newly developed methods for gearbox fault diagnosis or condition monitoring under variable working conditions.

Table 1 Comparison of several representative datasets

|  | BJTU WT-planetary gearbox dataset[1] | SEU planetary gearbox dataset[2] | UO variable speed bearing dataset[3] | MCC5-THU gearbox fault diagnosis datasets |
|---|---|---|---|---|
| Number of fault types | 5 | 5 | 3 | 7 |
| Number of vibration signals | 4 | 8 | 6 | 6 |
| Sampling frequency | 48kHz | 5.12kHz | 200kHz | 12.8 kHz |
| Sampling period | 5min | —— | 10s | 60s |
| Number of steady conditions | 8 | 2 | 0 | 24 |
| Key variables | Speed | Speed, Load | Speed | Speed, Load |
| Number of transitional conditions | —— | —— | 18 | 48 |
| Number of compound faults | —— | —— | —— | 2 |
| Number of fault degrees of severity | —— | —— | —— | 3 |

# BACKGROUND

The operating environment of gearboxes is both complex and harsh[4,5]. During the start-stop phase, the load, speed, lubrication conditions, and other operational parameters undergo changes, resulting in the gearbox functioning under time-varying speed and load conditions. Consequently, the variable working conditions (i.e., operating mode) of gearboxes lead to varying distributions of characteristics and frequencies for the same gear under different operating conditions. This variability, in turn, affects the robustness and accuracy of fault diagnosis models[6]. The significance of fault diagnosis algorithms lies in their capacity to efficiently identify and localize anomalies within complex systems, thereby facilitating timely and accurate remedial actions to ensure system reliability and performance optimization[7,8]. Therefore, it is of great significance to provide sufficient data support for studying fault diagnosis models under variable operating conditions.

# DATA DESCRIPTION

The datasets are collected based on different fault types, fault degrees, and working conditions, and contain a total of 240 sets of time series data. Each dataset contains 8 columns of data. The letters in the first row of each column indicate the meaning of the data in that column, as shown in Table 2. These datasets were collected based on a two-stage parallel gearbox under healthy and fault states. The internal structure diagram of the gearbox is shown in Figure 1. The data can be employed to evaluate the effectiveness of methods developed for gearbox fault diagnosis under time-varying speed or load conditions, such as the methods proposed in the literature [9-12].

Table 2 The meaning of each column in the dataset

| Letters | Meaning | Unit |
|---|---|---|
| speed | Motor output shaft key phase signal | dimensionless |
| torque | Gearbox input shaft torque | Nm |
| motor_vibration_x | Axial vibration acceleration of motor drive end | g |
| motor_vibration_y | Horizontal vibration acceleration of motor drive end | g |
| motor_vibration_z | Vertical vibration acceleration of motor drive end | g |
| gearbox_vibration_x | Axial vibration acceleration of gearbox intermediate shaft bearing seat | g |
| gearbox_vibration_y | Horizontal vibration acceleration of gearbox intermediate shaft bearing seat | g |
| gearbox_vibration_z | Vertical vibration acceleration of gearbox intermediate shaft bearing seat | g |

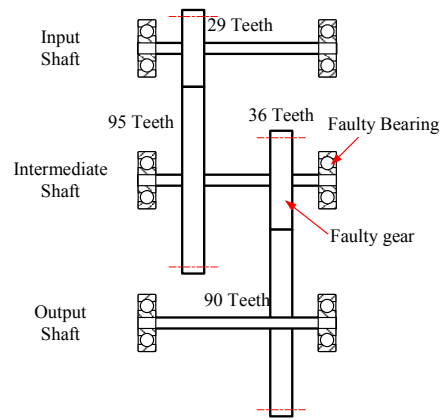

Figure 1 The internal structure diagram of the gearbox.

Furthermore, the gear module is 1.5, and the gear width is 10 mm. The 36-tooth gear on the intermediate shaft is a faulty gear, and the support bearing model ER16K at the end of the intermediate shaft near the 36-tooth gear is a faulty bearing. The specific parameters of the utilized ER16K bearing are reported in Table 3.

Table 3 The specific parameters of utilized ER16K bearing

| | |
|---|---|
| Inner diameter | 1 inch |
| Outer diameter | 2.0472 inch |
| Width | 0.749 inch |
| Ball diameter | 0.3125 inch |
| Number of balls | 9 |
| Pitch diameter | 1.516 inch |

Each dataset was measured with a sampling frequency of 12.8 kHz. The datasets were stored in the standard Excel format, ".csv," in a single column without a time stamp. They were collected at time-varying speeds or time-varying loads for a fixed duration of 60 seconds, with the set speed-time curve and load-time curve depicted in Figure 2(a) and Figure 2(b), respectively. The number of colors in the figure is used to distinguish groups of experiments. Taking the 0-2500-3000 situation marked in blue in Figure 2(a) as an example, it means that the speed is set to 3000 rpm within 10-20 seconds and within 40-50 seconds. At the same time, the rotation speed is set to 2500 rpm within 25-30 seconds.

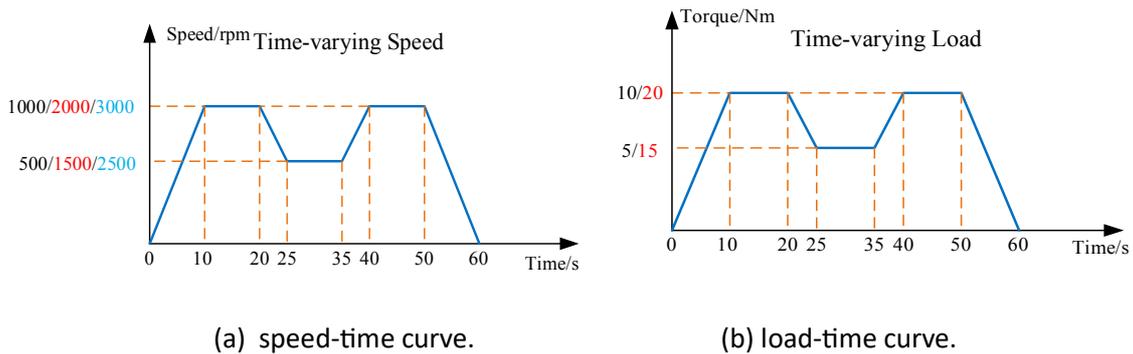

(a) speed-time curve.   (b) load-time curve.

Figure 2 The set time-varying curves.

During experiments under time-varying load conditions, the rotation speed was set to three constant values: 1000 rpm, 2000 rpm, and 3000 rpm. Similarly, during experiments under time-varying speed conditions, the load is set to two constant values: 10 Nm and 20 Nm.

The faults set in the experiment include both single gear faults and coincident gear and bearing faults. All faults were obtained by laser etching with a machining accuracy of 0.01mm. The details corresponding to the fault parameters for bearings and gears are reported in Table 4 and Table 5, respectively. In each scenario, in addition to the health status and "missing teeth" fault, three different fault degrees of severity are considered for the remaining four single faults ("gear wear," "gear pitting," "teeth crack," "teeth break") and two compound faults ("teeth break and bearing inner" and "teeth break and bearing outer"). Therefore, a total of 240 ".csv" files are included. In this context, a total of 24 steady conditions and 112 transitional condition combinations are considered.

Of note, the specific usage of the datasets is basically consistent with the usage of representative datasets. For researchers intending to leverage this dataset, we suggest constructing data encompassing the entire simulation lifecycle, reflecting real-world conditions, to evaluate the efficacy of their proposed methods. Additionally, researchers have the flexibility to generate tailored training and testing datasets as per their specific task demands and the contents of the provided .csv files.

Table 4 The details of bearing fault parameters

| Fault Parameter of Bearing | Fault Degree | | |
|---|---|---|---|
| | Light | Medium | High |
| Inner Raceway Fault Width | 0.1 mm | 0.3 mm | 0.5 mm |
| Outer Raceway Fault Width | 0.1 mm | 0.3 mm | 0.5 mm |

Table 5 The details of gear fault parameters

| Fault Parameter of Gear | Fault Degree | | |
|---|---|---|---|
| | Light | Medium | High |
| **Teeth Crack Depth** | 1/4 of the teeth height | 1/2 of the teeth height | 3/4 of the teeth height |
| **Gear Wear** | 1/3 of the teeth surface area | 1/2 of the teeth surface area | Full teeth surface area |
| **Teeth Break** | 1/4 of the teeth width | 1/2 of the teeth width | 3/4 of the teeth width |
| **Gear Pitting** | Fault diameter 0.5 mm | Fault diameter 1.0 mm | Fault diameter 1.5 mm |

Using the "miss_teeth_torque_circulation_3000rpm_20Nm" dataset as a case study, Figure 3 presents the visualization diagram. It is noteworthy that the rotational frequency of the intermediate shaft is determined to be 14.84Hz, while the meshing frequency of the small gear on the intermediate shaft is observed at 540.63Hz, accompanied by its sideband frequencies at 525.78Hz and 555.47Hz. These observations align with the typical characteristics associated with missing teeth faults.

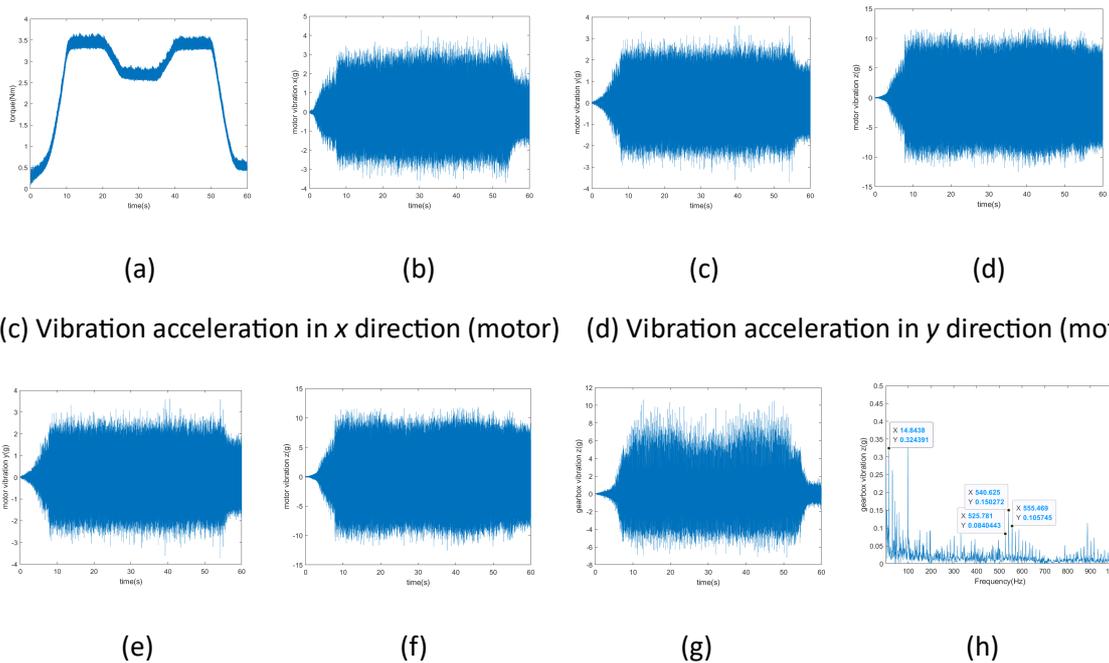

(a)　　　　　　　　(b)　　　　　　　　(c)　　　　　　　　(d)

(c) Vibration acceleration in *x* direction (motor)　(d) Vibration acceleration in *y* direction (motor)

(e)　　　　　　　　(f)　　　　　　　　(g)　　　　　　　　(h)

Figure 3 The visualization diagram of "miss_teeth_torque_circulation_3000rpm_20Nm" dataset. (a) represents the actual measured input shaft torque; (b-d) depict the vibration acceleration in the xyz directions of the motor; (e-g) showcase the vibration acceleration in the xyz directions of the gearbox's intermediate shaft; and (h) presents the envelope spectrum of the z-axis vibration acceleration of the gearbox's intermediate shaft.

In addition, it should be noted that due to the hysteresis of the motor and torque generator, the actual speed-time curve or load-time curve may differ slightly from the set curve. The torque load of the gearbox output shaft is applied by the magnetic powder brake, while the torque borne by the gearbox input shaft is measured by the torque sensor. In this context, the actual signal has been measured by the key phase sensor and torque sensor.

To better understand the setting conditions corresponding to the dataset files, the meanings of some file names are explained as Table 6.

Table 6 Examples of the dataset file descriptions

| Filename | Description |
| --- | --- |
| gear_pitting_H_speed_circulation_10Nm_1000rpm | a gearbox dataset for high teeth surface pitting. Single pit diameter is 1.5 mm. Gear output shaft torque is 10 Nm. The motor input shaft rotates at the 0-1000 rpm speed-time curve shown in Figure 2a. |
| gear_pitting_L_speed_circulation_20Nm_3000rpm | a gearbox dataset for light teeth surface pitting. Single pit diameter is 0.5 mm. Gear output shaft torque is 20 Nm. The motor input shaft rotates at the 0-3000 rpm speed-time curve shown in Figure 2a. |
| gear_pitting_H_torque_circulation-1000rpm_10Nm | a gearbox dataset for high teeth surface pitting. Single pit diameter is 1.5 mm. The motor input shaft rotates at a constant speed of 1000 rpm. The torque of the gearbox output shaft is between 0-10Nm, and the load-time curve is shown in Figure 2b. |
| gear_pitting_M_torque_circulation_3000rpm_20Nm | a gearbox dataset for medium teeth surface pitting. Single pit diameter is 0.5 mm. The motor input shaft rotates at a constant speed of 3000 rpm. The torque of the gearbox output shaft is between 0-20Nm, and the load-time curve is shown in Figure 2b. |
| health_speed_circulation_20Nm-3000rpm | a gearbox dataset for health gearbox. The motor input shaft rotates at a constant speed of 3000 rpm. Gearbox output shaft torque is 20 Nm. |
| miss_teeth_torque_circulation_1000rpm_10Nm | a gearbox dataset for missing a teeth gearbox. The motor input shaft rotates at a constant speed of 1000 rpm. The torque of the gearbox output shaft is between 0-10Nm, and the load-time curve is shown in Figure 2b. |
| teeth_break_H_torque_circulation_1000rpm_10Nm | a gearbox dataset for high teeth break. Three quarters of a tooth is broken. The motor input shaft rotates at a constant speed of 1000 rpm. The torque of the gearbox output shaft is between 0-10Nm, and the load-time curve is shown in Figure 2b. |
| gear_wear_M_torque_circulation_1000rpm_10Nm | a gearbox dataset for medium gear wear. 1/2 of the tooth surface is worn. The motor input shaft rotates at a constant speed of 1000 rpm. The torque of the gearbox output shaft is between 0-10Nm, and the load-time curve is shown in Figure 2b. |

| teeth_crack_L_torque_circulation_1000rpm_10Nm | a gearbox dataset for light teeth crack. Teeth crack depth reaches 1/3 of tooth height. The motor input shaft rotates at a constant speed of 1000 rpm. The torque of the gearbox output shaft is between 0-10Nm, and the load-time curve is shown in Figure 2b. |
|---|---|
| teeth_break_and_bearing_inner_H_torque_circulation_1000rpm_10Nm | a gearbox dataset for high tooth break and bearing inner ring compound failure. Three quarters of a tooth is broken. The fault width of the inner raceway is 0.5mm. The motor input shaft rotates at a constant speed of 1000 rpm. The torque of the gearbox output shaft is between 0-10Nm, and the load-time curve is shown in Figure 2b. |
| teeth_break_and_bearing_outer_H_torque_circulation_1000rpm_10Nm | a gearbox dataset for high tooth break and bearing outer ring compound failure. Three quarters of a tooth is broken. The fault width of the outer raceway is 0.5mm. The motor input shaft rotates at a constant speed of 1000 rpm. The torque of the gearbox output shaft is between 0-10Nm, and the load-time curve is shown in Figure 2b. |

## EXPERIMENTAL DESIGN, MATERIALS AND METHODS

The experimental setup for the dataset is shown in Figure 4, which consists of a 2.2 kW three-phase asynchronous motor, a torque sensor, a two-stage parallel gearbox, a magnetic powder brake acting as a torque generator, and a measurement and control system. This dataset aims to simulate and document various fault conditions of the 36-tooth gear on the intermediate shaft and its adjacent support bearings under different operating modes. The magnetic powder brake is used to apply a torque load to the gearbox. The actual torque endured by the gearbox input shaft can be measured by a torque sensor (model S2001, synthetic accuracy: $\pm 0.5\%$F.S). The speed sensor is used to measure the key phase signal of the motor output shaft, and the motor output shaft speed can be calculated from the key phase signal. The test rig, as depicted in Figure 4, is equipped with two three-axis vibration acceleration sensors (model TES001V, sensitivity: 100 mv/g) were used to measure motor and gearbox intermediate shaft triaxial vibrations along the x-, y-, and z-axes at a sampling frequency of 12.8 kHz. The datasets were collected and processed under 12 working conditions. In order to reduce the experimental error and measurement error caused by temperature, the temperature difference in the laboratory is controlled within the range of 2°C.

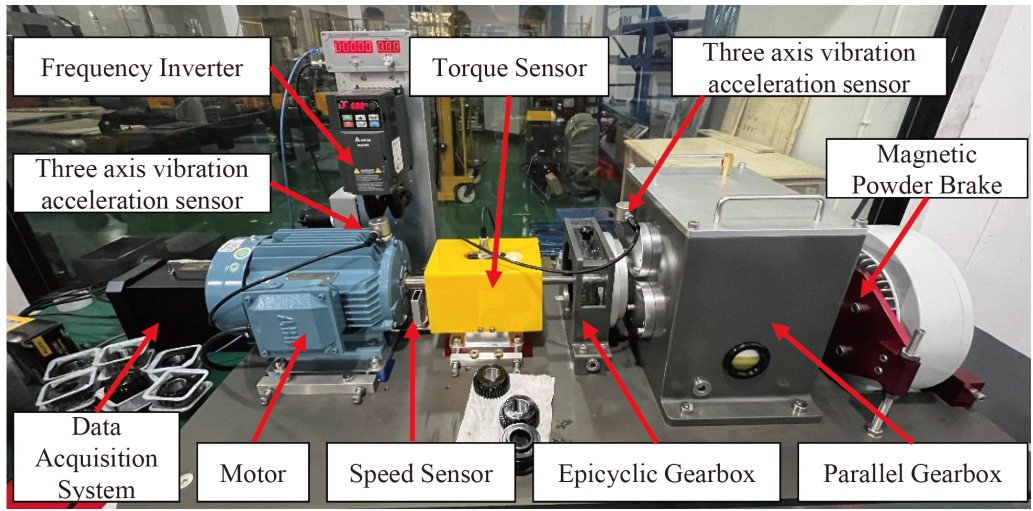

Figure 4 The actual Gearbox test rig.

# LIMITATIONS

It is important to mention that this paper includes a wide range of experiments covering different working conditions, types of faults, and levels of severity. However, due to limitations in our experimental setup, we haven't conducted experiments at very high speeds, such as 30000 rpm. We plan to address this in future versions of our research.

# ETHICS STATEMENT

*The authors have read and follow the ethical requirements for publication in Data in Brief and confirming that the current work does not involve human subjects, animal experiments, or any data collected from social media platforms.*

# CRediT AUTHOR STATEMENT

**Shijin Chen:** Experimental scheme design, Writing, Original draft preparation

**Zeyi Liu:** Conceptualization, Methodology, Writing- Reviewing and Editing

**Xiao He:** Conceptualization, Supervision

**Dongliang Zou:** Conceptualization, Supervision

**Donghua Zhou:** Conceptualization, Supervision

# ACKNOWLEDGEMENTS

*S. Chen and Z. Liu contributed equally to this paper. This work was supported by the China MCC5 Group PhD innovation project under grant MCC5PHD20231107, National Natural Science Foundation of China under grant 61733009, National Key Research and Development Program of China under grant 2022YFB25031103, and Huaneng Group Science and Technology Research Project under grant HNKJ22-H105.*

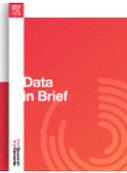

## DECLARATION OF COMPETING INTERESTS

The authors declare that they have no known competing financial interests or personal relationships that could have appeared to influence the work reported in this paper.

## REFERENCES

[1] D. Liu, L. Cui, W. Cheng. A review on deep learning in planetary gearbox health state recognition: Methods, applications, and dataset publication, *Measurement Science and Technology*, 2023, doi: DOI 10.1088/1361-6501/acf390.

[2] S. Shao, S. McAleer, R. Yan, P. Baldi. Highly accurate machine fault diagnosis using deep transfer learning, *IEEE Transactions on Industrial Informatics*, 2018, 15(4): 2446-2455.

[3] H. Huang, N. Baddour. Bearing vibration data collected under time-varying rotational speed conditions, *Data in Brief*, 2018, 21: 1745-1749.

[4] Y. Chen, M. Rao, K. Feng, G. Niu. Modified varying index coefficient autoregression model for representation of the nonstationary vibration from a planetary gearbox, *IEEE Transactions on Instrumentation and Measurement*, 2023, 72: 3511812.

[5] Y. Chen, M. Rao, K. Feng, M. Zuo. Physics-Informed LSTM hyperparameters selection for gearbox fault detection, *Mechanical Systems and Signal Processing*, 2022, 171: 108907.

[6] X. He, C. Li, Z. Liu, A real-time adaptive fault diagnosis scheme for dynamic systems with performance degradation, *IEEE Transactions on Reliability*, 2023, doi: 10.1109/TR.2023.3324539.

[7] T. Han, W. Xie, Z. Pei. Semi-supervised adversarial discriminative learning approach for intelligent fault diagnosis of wind turbine, *Information Sciences*, 2023, 648: 119496.

[8] J. Yao, T. Han. Data-driven lithium-ion batteries capacity estimation based on deep transfer learning using partial segment of charging/discharging data, *Energy*, 2023, 271: 127033.

[9] Z. Liu, C. Li and X. He, Evidential ensemble preference-guided learning approach for real-time multimode fault diagnosis, *IEEE Transactions on Industrial Informatics*, 2024, 20(4): 5495-5504.

[10] Y. Wei, Z. Xiao, S. Liu, K. Schröder, H. Peng, A. Sarr, X. Gu, A novel data augmentation and composite multiscale network for mechanical fault diagnosis, *IEEE Transactions on Instrumentation and Measurement*, vol. 72, pp. 1-12, 2023, Art no. 3525912.

[11] Z. Liu, J. Zhang, X. He, A discrimination-guided active learning method with marginal representation for industrial compound fault diagnosis, *IEEE Transactions on Automation Science and Engineering*, 2023. doi: 10.1109/TASE.2023.3325271.

[12] Z. Liu, J. Zhang, X. He, Q. Zhang, G. Sun, D.-H. Zhou, Fault diagnosis of rotating machinery with limited expert interaction: a multi-criteria active learning approach based on broad learning system, *IEEE Transactions on Control Systems Technology*, vol. 31, no. 2, pp. 953-960, 2023.